\documentclass{cpcauth}
\usepackage{psfig}
\newcommand{\D}{{\rm d}}
\begin{document}
\begin{frontmatter}
\title{Anisotropic scaling and generalized conformal invariance 
at Lifshitz points}

\author[NA]{Malte Henkel\thanksref{ca}} and
\author[NA,ER]{Michel Pleimling}
\address[NA]{Laboratoire de Physique des Mat\'eriaux$^{\, 2}$, 
Universit\'e Henri Poincar\'e Nancy I, B.P. 239,\\
F--54506 Vand{\oe}uvre--l\`es--Nancy Cedex, France}
\address[ER]{Institut f\"ur Theoretische Physik 1, 
Universit\"at Erlangen-N\"urnberg, D--91058 Erlangen, Germany}
\thanks[ca]{Corresponding author:
{\tt henkel@lpm.u-nancy.fr}}
\thanks[an]{Laboratoire associ\'e au CNRS UMR 7556}
 
\begin{abstract}
A new variant of the Wolff cluster algorithm is proposed for simulating systems
with competing interactions. This method is used in a high-precision study 
of the Lifshitz point of the $3D$ ANNNI model. At the Lifshitz point, several 
critical exponents are found and the anisotropic scaling of the correlators is 
verified. The functional form of the two-point correlators is shown to be 
consistent with the predictions of generalized conformal invariance. 
\end{abstract}
\end{frontmatter}

Competing interactions are encountered in a large variety of physical 
systems such as, among others, magnets, alloys or ferroelectrics. 
These interactions may lead to
rich phase diagrams with a multitude of phases as 
well as to special multicritical
points called Lifshitz points (LP). At an LP, a disordered, a uniformly
ordered and a periodically ordered phase become indistinguishable \cite{Hor75}.
A large number of systems (magnets, ferroelectric liquid crystals, uniaxial
ferroelectrics, block copolymers) have been shown to posses an LP.

This kind of systems may be mimicked by
spin models with competing interactions.
The simplest of these models is the well-known ANNNI (axial next
nearest neighbour Ising) model \cite{Sel88,Neu98}. 
Because of the presence of the competing interactions, previous 
Monte Carlo studies of the transition(s) from the disordered 
high-temperature phase to the ordered low-temperature phases
exclusively used single-spin flip algorithms \cite{Sel78,Kas85}.
Due to the critical slowing-down inherent to these methods, 
those numerical studies were limited to rather small system sizes.
However, large systems are needed for the precise computation 
of critical quantities, such as  
critical exponents or critical correlators.

In this contribution, we present a new variant of the Wolff cluster algorithm
\cite{Wol89} specifically designed for the simulation of systems with 
competing interactions near criticality. Similarly, we also generalize
the recently proposed method of Evertz and von der Linden 
for the computation of correlation functions of infinite systems \cite{Eve01}
to systems with competing interactions. 
As an example, we present results obtained
{\em at} the LP of the three-dimensional ANNNI model, 
and we shall concentrate on the LP critical exponents and critical correlators.
In particular, our data for the scaling of the two-point correlation functions
are consistent with the theoretical predictions of generalized conformal 
invariance \cite{Hen97,Ple01}.

Problems coming from critical slowing-down encountered when 
using local Monte-Carlo dynamics are alleviated by using non-local methods,
such as the Wolff cluster algorithm \cite{Wol89}.
For the Ising model with only a nearest-neighbour coupling $J$,
this algorithm may be described as follows: 
one chooses randomly
a lattice site, the seed, and then builds up iteratively a cluster. 
If $i$ is a cluster site (with spin $s_i$) a site $j$ neighbouring the
cluster site $i$ will be included into the cluster with probability 
$p= \frac{1}{2} \left( 1 + \mbox{sign} \left(
s_i s_j \right) \right) \left( 1 - \exp \left[ - 2 J /(k_B T) \right] \right)$.
One ends up with a cluster of spins having all the same sign, see Fig. 1a,
and which is flipped as a whole. 
This kind of same-sign clusters is 
obviously not adapted to our problem because of the competing interactions.

\begin{figure}
\centerline{\psfig{figure=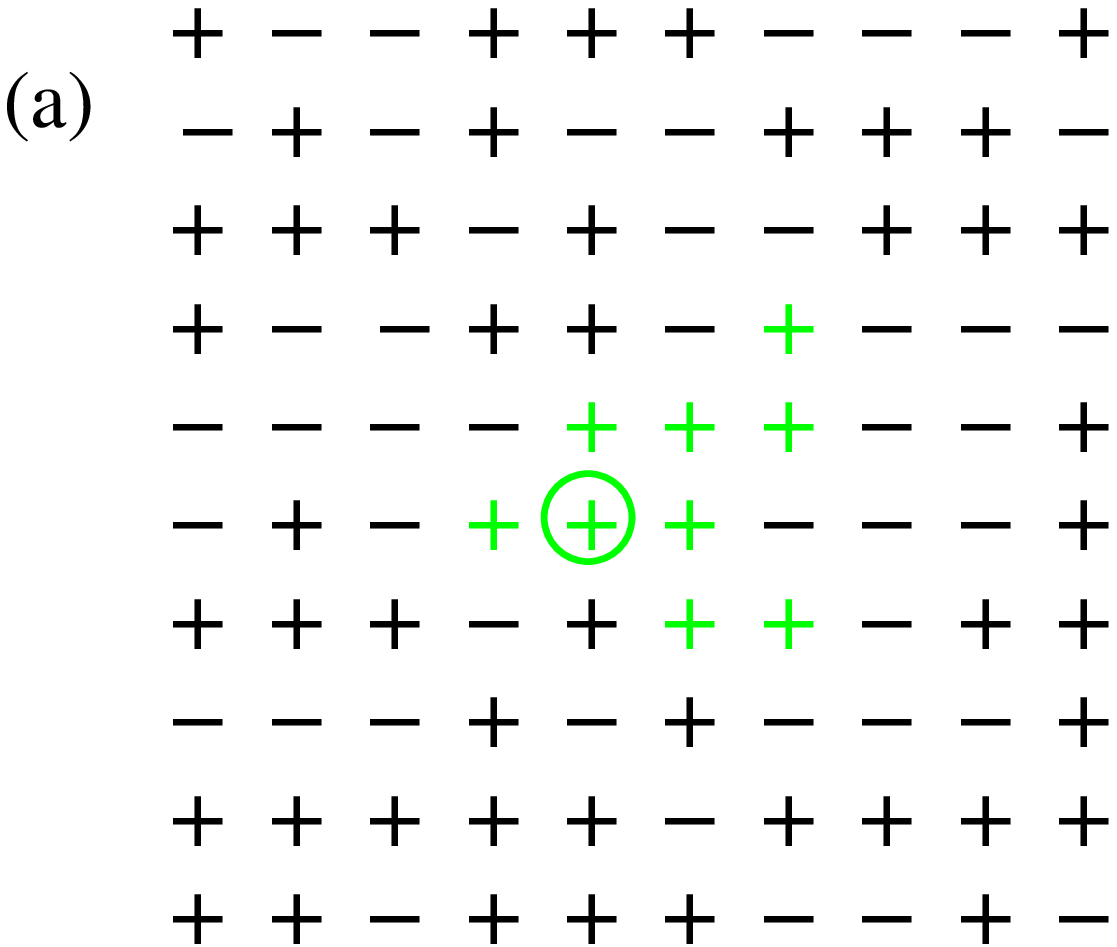,width=7cm}}
\vspace*{1cm}
\centerline{\psfig{figure=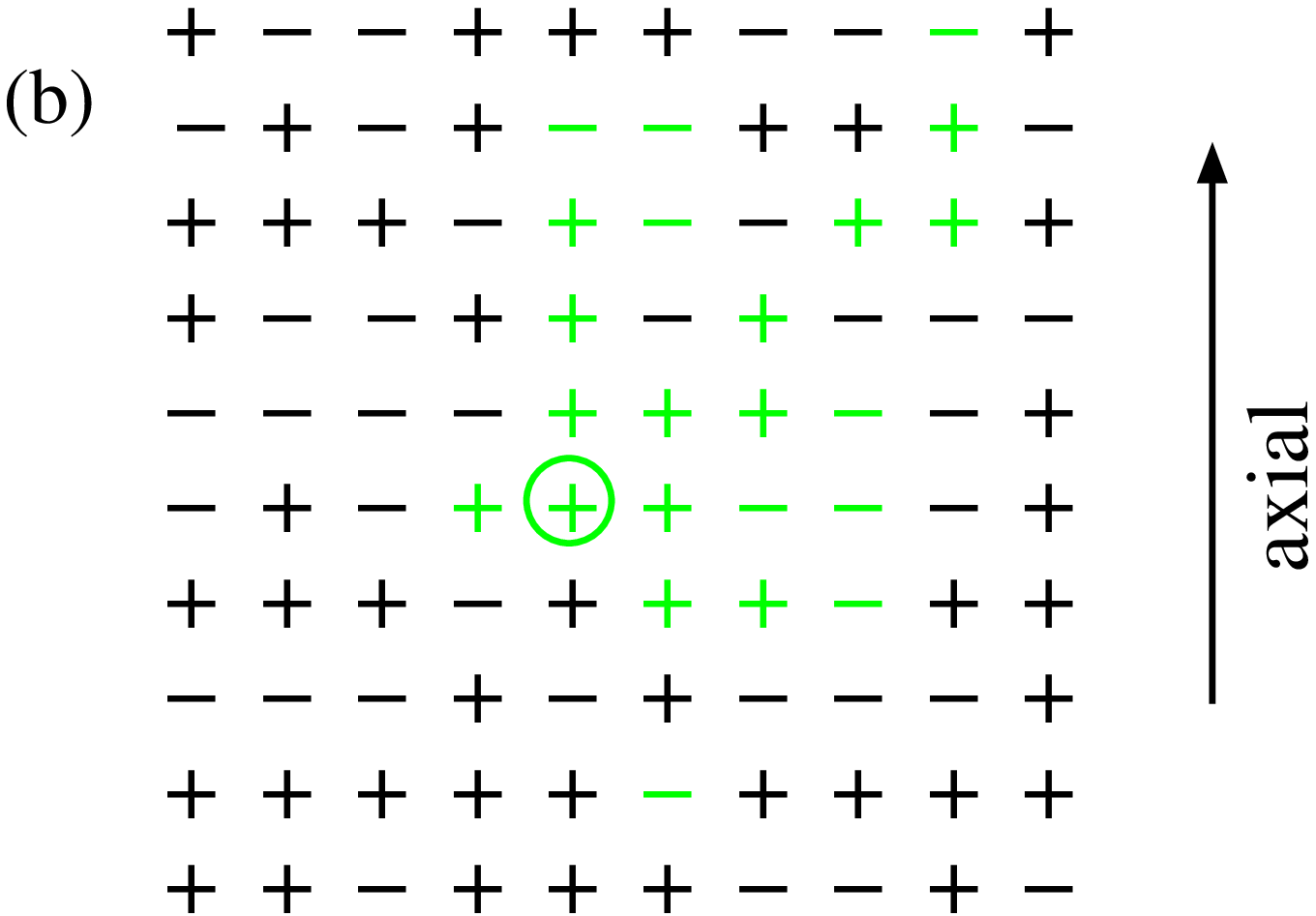,width=7cm}} 
\caption{Typical clusters (gray spins) obtained (a) by the Wolff algorithm
and (b) by our modified algorithm for systems with competing interactions,
here shown in two dimensions for simplicity.
The competition takes place in the axial direction. The spin enclosed by
a circle is the seed.}
\label{fig1} \end{figure}

The modified cluster algorithm we propose for simulating spin models with
competing interactions is in the following presented for the ANNNI model.
A generalization to other models is straightforward.
The Hamiltonian of the $3D$ ANNNI model is, with $s_{xyz} = \pm 1$,
\begin{eqnarray}
H &=& - J \sum\limits_{xyz} s_{xyz} \left( s_{(x+1)yz} + s_{x(y+1)z} +
s_{xy(z+1)} \right) \nonumber \\
& &+ \kappa \, J \sum\limits_{xyz} s_{xyz}s_{xy(z+2)} \label{Gl:annni}
\end{eqnarray}
whereas $J >0$ and $\kappa > 0$ are coupling constants. 
In the axial $z$-direction competition between ferromagnetic
nearest-neighbour and antiferromagnetic next-nearest-neighbour couplings
occurs, leading to a rich phase diagram \cite{Sel92}.

Our proposed modified cluster algorithm starts 
with a randomly chosen seed and builds up iteratively
a cluster. Consider a newly added cluster lattice site $i$ with spin $s_i$.
A lattice site $j$ with spin $s_j$ nearest neighbour to $i$ is included
with probability $p_{\rm n}=p$, whereas an {\it axial} next-nearest
neighbour site $k$ with spin $s_k$ is included with probability
$p_{\rm a}=\frac{1}{2} \left( 1 - \mbox{sign} \left(
s_i s_k \right) \right) \left( 1 - \exp \left[ -  2 J \kappa 
/(k_B T) \right] \right)$. Thus the final cluster, which will be flipped
as a whole, contains spins of both signs, as shown in Fig. 1b. 
At variance with the traditional Wolff method, the flipped spins are not
necessarily connected by nearest-neighbour couplings. Ergodicity and detailed
balance are proven as usual. This algorithm works extremely
well in the paramagnetic phase, in the ferromagnetic phase and in the 
vicinity of the LP.
It has not yet been subjected to stringent tests in the modulated region 
where large free energy barriers make simulations very difficult 
(we point out recent progress achieved in microcanonical 
simulations of finite ANNNI systems \cite{Was01}). 
Note that our algorithm differs from the same-sign cluster algorithm 
proposed by Luijten and Bl\"{o}te \cite{Lui95} for the simulation of 
systems with long-range interactions.

In an interesting work, Evertz and von der Linden \cite{Eve01} 
have proposed a new numerical scheme, based on the Wolff cluster algorithm, 
for the direct computation of two-point functions of an infinite lattice. 
In their approach, each cluster is started at the same lattice site and not 
from a randomly chosen site. After a certain
number of cluster flips, the two-point functions can be measured 
within a certain region, this region getting larger with increasing number of 
flips. With this method, correlation functions of the infinite lattice are 
obtained at temperatures above the critical temperature \cite{Eve01}. 
At the critical temperature, finite-size effects are greatly
reduced as compared to more traditional approaches. 
We have generalized the method of \cite{Eve01} to systems with competing 
interactions by replacing the usual Wolff clusters
with our modified clusters. This enables us to compute with unprecedented 
precision two-point correlators in spin systems
with competing interactions.

In the following, we report results of a large-scale Monte Carlo study
of the uniaxial LP encountered in the $3D$ ANNNI model where the
new algorithms introduced above have been used with great success \cite{Ple01}. 
The whole study took the equivalent of more than two CPU-years
on a DEC alpha workstation. For the investigation
of LP properties a precise location of this point is mandatory.
One possibility to locate this point is given by the structure factor 
$S(q)=\sum\limits_{\bf{r}}\langle S_{\bf{0}}\,S_{\bf{r}}\rangle
\,\exp (\mbox{i} q z)$
as the transition between uniformly ordered and periodically ordered 
phases shows up as a shift of the maximum of $S(q)$ from $q=0$ 
to a non-zero value \cite{Sel78}.
Having found a reliable value for $\kappa_L$, the LP critical temperature
is then obtained by standard methods. In Table 1, 
we compare the location of the 
LP as obtained from our data with previous estimations. Note that 
Kaski and Selke \cite{Kas85} did not attempt an 
independent determination of $\kappa_L$
but merely determined $T_L$ for a $\kappa$ in close vicinity to $\kappa_L$ as
obtained from a high-temperature (HT) series expansion \cite{Oit85}. 
{}From Table 1 the increase in precision is evident.

\begin{table}
\caption{Estimated location of the LP in the $3D$ ANNNI
model defined on a cubic lattice, as obtained from high-temperature
expansions (HT) and Monte Carlo simulations (MC).}
\vspace*{0.2cm}
\begin{tabular}{|c|c|c|}
\hline
& $\kappa_L$ & $T_L$ \\
\hline 
HT \cite{Oit85} & $0.270 \pm 0.005$ & $3.73 \pm 0.03$ \\
\hline
MC \cite{Kas85} & $0.265$ & $3.77 \pm 0.02$ \\
\hline
present work & $0.270 \pm 0.004$ & $3.7475 \pm 0.005$ \\
\hline
\end{tabular}
\end{table} 

The uniaxial LP is a strong anisotropic critical point where the
correlation lengths parallel and perpendicular to the axial direction diverge
with different critical exponents: 
$\xi_{\|,\perp} \sim | T- T_L|^{-\nu_{\|,\perp}}$. The value of the anisotropy
exponent $\theta=\nu_\|/\nu_\perp$ is in good approximation 
equal to 1/2 \cite{Die00}.
In order to take into account the special finite-size effects coming from the
anisotropic scaling at the LP \cite{Bin89}, large systems of
anisotropic shape with $L \times L \times N$ spins with $20 \leq L \leq 240$
and $10 \leq N \leq 100$ have been simulated.

Our estimates for the LP exponents $\alpha,\beta,\gamma$ are given in Table 2.
For the first time, these ANNNI model exponents are computed independently. 
They were found by investigating effective exponents which yield the
critical exponents in the limit $T \longrightarrow T_L$ \cite{Ple01,Ple98} 
(provided that finite-size effects can be neglected).
Our error bars take into account both the
sample averaging and the uncertainty in the location of the LP.
The reliability of our data is illustrated by the agreement of the 
independently estimated exponents $\alpha$, $\beta$ and $\gamma$ with the 
scaling relation $\alpha + 2 \beta + \gamma = 2$ up to $\approx0.8\%$. 
Furthermore, the agreement with a recent two-loop calculation \cite{Die01} 
is remarkable. 

\begin{table}
\caption{Critical exponents at the LP of the $3D$ ANNNI model,
as obtained from Monte Carlo simulations (MC) 
and renormalized field theory (FT).
The numbers in brackets give the estimated error in the last digit(s).}
\vspace*{0.2cm}
\begin{tabular}{|c|c|c|c|c|c|}
\hline
& $\alpha$ & $\beta$ & $\gamma$ & $(2-\alpha )/\gamma$ & $\beta/\gamma$ \\
\hline
MC \cite{Kas85} & & $0.19(2)$ & $1.40(6)$ & & $0.14(2)$ \\
\hline
FT \cite{Die00} & & & & 1.27& 0.134 \\
\hline
FT \cite{Die01} & 0.160 & 0.220 & 1.399 & 1.315 & 0.157 \\ \hline
present work & $0.18(2)$ & $0.238(5)$ & $1.36(3)$ 
& $1.34(5)$ & $0.175(8)$\\
\hline
\end{tabular}
\end{table}

Fig. 2 summarizes our data obtained for the LP spin-spin correlator 
$C({\bf r}_\perp, r_{\|} ) = \langle S_{{\bf r}_\perp, r_{\|}} \,
S_{\bf{0}_\perp, 0} \rangle$.
It shows the scaling function $\Phi(u)$ related to the correlator by
\begin{equation}
C({\bf r}_\perp, r_{\|} )= r_\|^{-2x/\theta} \, \Phi(r_\perp^\theta/r_\|)
\end{equation} 
and the scaling dimension $x=(2 + \theta)/(2 + \gamma/\beta)$
\cite{Hen97,Ple01}. In this analysis, we use $\theta= 1/2$, since
the small deviations from this value obtained in recent field-theoretical 
calculations \cite{Die00,Die01} are not yet distinguishable from the 
purely numerical errors in our data. The data shown in Fig. 2 have been 
obtained after more than five million cluster updates for a system with 
$200 \times 200\times 100$ spins. They permit a nice visual test of the data 
collapse and establish scaling, for the first time also at the LP.

\begin{figure}
\centerline{\psfig{figure=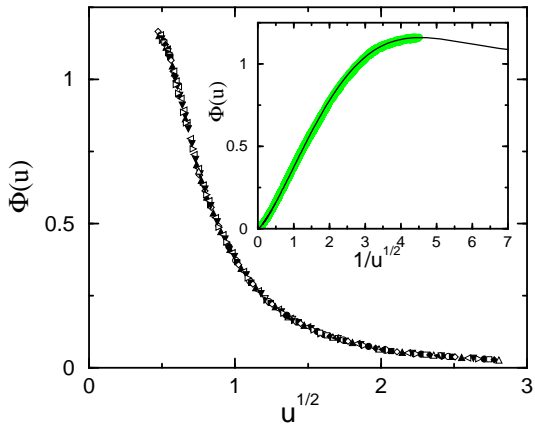,width=7cm}}
\caption{Scaling funtion $\Phi(u)$ versus 
$u^{1/2}=(\sqrt{r_\perp}/{r_\parallel})^{1/2}$ for $\kappa=0.270$ 
and $T=3.7475$. The different symbols correspond to different 
values of $r_\perp$. Inset: comparison of the full data set of 
$1.7 \times 10^4$ points for the scaling function
$\Phi(u)$ (gray points) with the analytical prediction 
resulting from reference \cite{Hen97} (full curve).
\label{fig2}} 
\end{figure}

The inset of Fig. 2 is a direct comparison of our numerical
data with the theoretical prediction resulting from a generalization of
conformal invariance to strong anisotropic criticality \cite{Hen97}.
Applied to the LP of the $3D$ ANNNI model and assuming $\theta=1/2$, 
the scaling function $\Omega(v)= v^{-2x/\theta} \, \Phi(1/v)$ should
satisfy the differential equation
\begin{equation} \label{Gl:DiffGl}
\alpha_1 \frac{\D^{3} \Omega(v)}{\D v^{3}} - v^2 \frac{\D \Omega(v)}{\D v} 
- \frac{2x}{\theta} v \Omega(v) = 0
\end{equation}
where $\alpha_1$ is a constant. It turns out \cite{Hen97,Ple01} that the 
functional form of $\Phi(u)$ only depends on a single universal parameter $p$ 
which can be determined from the Monte Carlo data by analyzing ratios of 
moments of $\Phi$. As shown in \cite{Ple01}, different moment ratios yield
consistent values for this parameter. The mean value of $p$ 
obtained in this way has been used
in the inset of Fig. 2. The nice agreement between theory and numerical data 
provides strong evidence for the applicability of the generalized conformal 
invariance to the strongly anisotropic critical behaviour realized at LPs. 
A similar agreement between numerics and
theory has been obtained for the energy-energy correlation function.

In conclusion, we have presented a new variant of the Wolff cluster algorithm 
designed for the study of systems with competing interactions. Applied to
the LP of the $3D$ ANNNI model, the anisotropic scaling of the two-point
correlators could be confirmed directly for the first time. The form of the
scaling function is in agreement with the hypothesis of generalized conformal
invariance.

\end{document}